\documentclass[twocolumn,pre]{revtex4}

\usepackage{natbib}
\usepackage{multirow}
\usepackage{amsmath}
\usepackage{graphicx}

\newcommand{\defn}{\textbf}

\newcommand{\eref}[1]{(\ref{#1})}
\newcommand{\av}[1]{\langle#1\rangle}
\newcommand{\al}{\alpha}
\newcommand{\T}{\Theta}
\newcommand{\Ta}{\T_{ij}}
\newcommand{\Taa}{\T_{jm}}
\newcommand{\Tb}{\T_{km}}
\newcommand{\Tc}{\T_{lq}}
\newcommand{\e}{\mathrm{e}}

\newcommand{\Ord}{\mathrm{O}}

\newcommand{\set}[1]{\{#1\}}

\newcommand{\half}{{1\over2}}
\newcommand{\etal}{\textit{et al.}}
\newcommand{\mean}[1]{\overline{#1}}
\newcommand{\cA}{A}

\newcommand{\cF}{\psi}

\newcommand{\cS}{\mathcal{S}}

\newcommand{\matA}{\mathbf{A}}

\newcommand{\bda}{\vec{\alpha}}
\newcommand{\bds}{\mathbf{s}}
\newcommand{\bdr}{\mathbf{r}}

\newcommand{\ex}[1]{\e^{#1}}
\newcommand{\Ns}{\textit{N}-space}
\newcommand{\ns}{\textit{n}-space}

\newlength{\figurewidth}
\ifdim\columnwidth<20cm
\else
\fi
\begin{document}
\title{On Minimum Violations Ranking in Paired Comparisons}
\author{Juyong Park}
\affiliation{Department of Physics and Center for the Study of Complex Systems, The University of Michigan, Ann Arbor, Michigan 48109}
\begin{abstract}
Ranking a set of objects from the most dominant one to the least, based on the results of paired comparisons, proves to be useful in many contexts.  Using the rankings of teams or individuals players in sports to seed tournaments is an example.  The quality of a ranking is often evaluated by the number of violations, cases in which an object is ranked lower than another that it has dominated in a comparison, that it contains.  A minimum violations ranking (MVR) method, as its name suggests, searches specifically for rankings that have the minimum possible number of violations which may or may not be zero.  In this paper, we present a method based on statistical physics that overcomes conceptual and practical difficulties faced by earlier studies of the problem.
\end{abstract}
\maketitle

\section{Introduction}
A paired comparison of a set of $n$ objects $\set{O_1,O_2,\ldots,O_n}$ is conducted by forming a pair of objects from the set and then deciding the better (stronger, preferred, more dominant, etc., depending on the context) one between the pair.  If all ${n\choose 2}$ possible pairs are compared, the comparison is called \emph{complete}, and \emph{incomplete} otherwise.  In either case, it is commonplace to attempt to rank the objects from the ``best'' (ranked highest, at first place) to the ``worst'' (ranked lowest, at $n$-th place).  An example is competitive sports. In a typical sports match, two teams or individuals play against each other, and a ``winner'' and a ``loser'' are pronounced at its conclusion.  It is well known that many sports, including soccer, tennis, and American collegiate football, maintain an official ranking of teams or players based on their performances in those matches.  These rankings are then often used to seed tournaments~\cite{Stefani:1997bb}.

We can find examples in other fields as well. In biology, acts of aggression or submission of an animal towards another motivates the study of ``dominance hierarchies'' within the herds or communities of the animals~\cite{Cole:1981hy,Lott:1979hg}.  Another instructive example is a certain form of a decision-making process.  Imagine a city council that, due to a limited budget, can only approve a handful among all the development projects that its constituents are demanding (improving schools, building new hospitals, repairing highways, and so forth).  Assume that the council has conducted paired comparisons of the projects so that it is known which the constituents prefer, given two projects at a time~\footnote{A paired comparison is known to be effective when people are asked to show a preference of one alternative over another~\cite{Goddard:1983sf}}.  What would be the optimal selection of projects to pursue? If the council is able to rank the projects in such a way that no project is ranked lower than another project to which it was actually preferred to, it may just approve the projects starting from the top, possibly escaping the blame for ``messing up'' priorities.  However, this may not always be possible because, in paired comparisons, preferences are not necessarily \emph{transitive}.  Letting $P_1>P_2$ mean that ``Project $P_1$ is preferred to $P_2$''~\footnote{Note that most concepts introduced here translate naturally to other contexts (sports, animals, etc.). For example, $T_1 > T_2$ in sports would imply ``team $T_1$ has beaten team $T_2$ (in a match)''.}, we could certainly imagine a \emph{cycle} of preferences, e.g. school $>$ hospital, hospital $>$ bridge, and bridge $>$ school. (The reason for this is that the criteria of comparison may depend on the specific objects being compared~\cite{Kendall:1955bw}.)  If there is such a cycle, any ranking will inevitably show at least one instance of a project being ranked lower than another that it has been preferred to.  This is called a \defn{violation} or an \defn{inconsistency} of a ranking~\cite{Ali:1986ey}.  The next wisest strategy, then, would be to come up with a ranking that has the minimum number of violations, which is called a \defn{minimum violations ranking (MVR)} of the paired comparison.  Since the number of violations cannot be negative, a zero-violation ranking is a special case of MVR.

\begin{figure}[t]
\includegraphics[width=60mm]{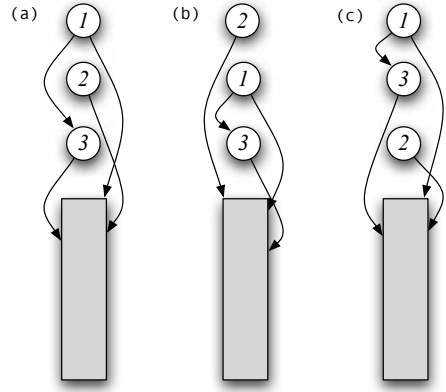}
\caption{An illustration of the multiplicity of minimum violations rankings.  Three objects $(1,2,3)$ are depicted as labeled nodes, and the rest of the system as a gray box.  An arrow points from the winner to the loser in a paired comparison.  In (a) and (b), we see that two unbeaten objects may change places freely at the top, without affecting the number of violations.  However, being unbeaten does not guarantee being at the top, as we see in (c) which has the same number of violations as (a) and (b) have.}
\label{multiplicity}
\end{figure}

Therefore, the advantage of having an MVR is clear (i.e., it is the best that one can do to minimize messed-up priorities), with its meaning very well-defined.  Practically, however, the application of this method to real situations has been limited for a couple of reasons.  First, the exact algorithm that finds an MVR of a paired comparison is slow ($\Ord(n^4)$)~\cite{Ali:1986ey}. This is reflected in the fact that, despite a century-long obsession of dedicated fans with figuring out the ``best'' football school, only recently has an MVR been found by Coleman for the American college football schedule that is composed of more than a hundred teams~\cite{Coleman:2005vm}.  Some of the previous studies on MVR have tried to introduce some heuristic methods to speed up the search, but have had to make sacrifices in the final results~\cite{Ali:1986ey,Kendall:1962zq,Kendall:1955bw,Vries:1998no}.  Second, the issue of the \emph{multiplicity} of MVRs has been ignored in these studies.  For example, if there are two ``unbeaten'' objects (i.e. those not have lost in any comparison), then it cannot matter to us which one is placed first and which second, since both cases will give the same number of violations.  This is shown schematically in Fig.~\ref{multiplicity}~(a)~and~(b). In the figure, three objects $(1,2,3)$ are drawn as labeled nodes that sit at the top three spots, and an arrow is drawn from the winner to the loser of a paired comparison.  To make matters more complex, being unbeaten may even not guarantee being at the top among other unbeaten ones, as we see in Fig.~\ref{multiplicity}~(c).

This raises the question of fairness: going back to our example of a city council, there can now arise a situation where $P_1$ should get approved over $P_2$ according to one MVR, but the opposite is true according to another MVR.  The city council is back in a difficult position of having to decide on one, despite the fact that both MVRs are equally qualified.  Is there a way to break out of this situation, without adding another \emph{ad hoc} criterion?  Here we propose a natural solution, which is \emph{to treat all MVRs as equally probable, and take the mean of each object's ranks}.  We then arrange the objects in the increasing order of these mean ranks to obtain a final standing.  For example, assume that the three rankings shown in Fig.~\ref{multiplicity} are the only possible MVRs of a some paired comparison.  Then the mean ranks of the three objects $(1,2,3)$ are $\set{1+2+1,2+1+3,3+3+2}/3=\set{1.33,2,2.67}$, from which we obtain the final standings $\set{1,2,3}$.  Although this looks identical to Fig.~\ref{multiplicity}(a), they must not be confused --- Fig.~\ref{multiplicity}(a) merely represents one specific MVR (among many) of the paired comparison.

While being straightforward and fair to all MVRs as requested, this solution seems to ask us to find all possible MVRs.  Assume that $m$ pairs have been compared. Since there are $n!$ different rankings of $n$ objects, there will be on average $n!/m$ rankings that show a particular number of violations.  This can be a huge number, even for a common data set: in $2004$, the American college football schedule consisted of $117$ schools playing $622$ games, which gives $117!/622\simeq10^{190}$!

An exhaustive search through such a large state-space is therefore clearly impossible, and so we are in need of some sampling technique. As a matter of fact, we find ourselves in a very familiar territory: we are trying to average certain observables (ranks of objects) over a set of a large number of microstates (rankings), given a constraint (number of violations).  These are the features of a typical combinatoric problem that can be formulated as a statistical physics model~\cite{Nishimori:2001ib,Mertens:1998sd,Sasamoto:2001yp,Ferreira:1998jq}, and that is precisely what we will do in the following section.

\section{Definition of the Model}
We represent the results of a paired comparison of the set of $n$ objects as an $n\times n$ matrix $\matA$, called the \defn{adjacency matrix}.  Its element $\cA_{ij}$ is equal to $1$ if $j$ has beaten $i$ in a comparison and $0$ otherwise.  In an \emph{incomplete} comparison, there exist pairs of objects for which $\cA_{ij}=\cA_{ji}=0$.  However, we require that the set of objects cannot be divided into two sets $S_1$ and $S_2$ such that $A_{ij}=0$ and $A_{ji}=0$ for all $O_i\in S_1$ and $O_j\in S_2$. This is necessary because if such a division is possible, objects in $S_1$ and $S_2$ are completely independent, and therefore it would be nonsensical to try to rank them on a single linear scale.

\subsection{Expansion of the State Space}

\begin{figure}[t]
\includegraphics[width=80mm]{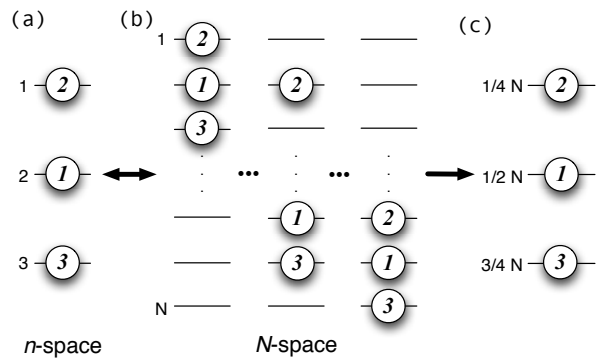}
\caption{There exists a natural mapping between (a) the $n$-space and (b) the $N$-space.  Each ranking in $n$-space is mapped to ${N\choose n}$ rankings with identical relative standings in the $N$-space. The ranks in the $n$-space are linearly scaled into (c) the mean ranks in the $N$-space according to Eq.~\eref{scaling}}
\label{mapping}
\end{figure}

The definition of a ranking of $n$ objects we have used so far is an ordering of the $n$ objects from the first place to the last ($n$-th).  More precisely, we assign to each object an integer \emph{rank} $r_i$ ($1\le r_i\le n$), such that the \emph{ranking} $\bdr=\set{r_1,r_2,\dots,r_n}$ is a permutation of $\set{1,\ldots,n}$. (Although we do not allow ties in a ranking, they can be produced in the final standings if the mean ranks of multiple objects are equal.)  We can also visualize this as $n$ objects occupying $n$ ordered slots, which we will call the \defn{\textit{n}-space}. This is illustrated in Fig.~\ref{mapping}(a) for $n=3$.  It will be shown later that it is more convenient to work in a larger \defn{\textit{N}-space}, where the same $n$ objects occupy $n$ slots among $N$ ($N\gg n$) ordered slots.  The ranking in the \textit{N}-space is similarly defined as $\bds=\set{s_1,s_2,\ldots,s_n}$, where $s_i$ is the label of the slot occupied by object $i$. This is illustrated in Fig.~\ref{mapping}(b).

We can define a natural mapping between a ranking in the $n$-space and rankings in the \Ns~that have the same relative standings of the objects (i.e., in terms of being higher or lower than others), hence the same number of violations.  Under this mapping, each ranking in the $n$-space corresponds to exactly ${N\choose n} =  N^n/n!$ rankings in the $N$-space.  Therefore, Fig.~\ref{mapping}(a) gets mapped onto Fig.~\ref{mapping}(b).  

Now, consider an object ranked $r$ in some ranking in the \ns.  Its mean rank $\tilde{s}(r)$, averaged over the mapping of the ranking in the \Ns, is given by
\begin{eqnarray}
\tilde{s}(r) &=& \sum_{x=1}^{N}x\cdot\frac{x^{r-1}}{(r-1)!}\cdot\frac{(N-x+1)^{n-r}}{(n-r)!}\biggl/\biggl(\frac{N^n}{n!}\biggr) \nonumber \\
%          &=& \sum_{i=1}^N \frac{i^r}{(r-1)!}\frac{(N-i+1)^{n-r}}{(n-r)!}\biggl/\biggl(\frac{N^n}{n!}\biggr) \nonumber \\
%          &=& \frac{r}{N^n}{n\choose r}\sum_{i=1}^N i^r(N-i+1)^{n-r} \nonumber \\
          &\equiv& r\frac{N}{n+1}\mathcal{L}(N,n,r).
\label{scaling}
\end{eqnarray}

Numerical studies strongly suggest that $\mathcal{L}(N,n,r) = 1+\Ord(n/N) \to 1$ for $N\gg n$.  If this is true (and for all the examples that we will be presenting in the final part of this paper, it is), Eq.~\eref{scaling} tells us that $\tilde{s}(r)$ is identical to $r$ up to a constant prefactor, i.e. $\tilde{s}(r)=r\cdot(N/n+1)\equiv r\cdot C$.  Assume that we have a set $R=\set{\bdr^a,\bdr^b,\ldots}$ of rankings in the \ns~(e.g., of all MVRs), and that we are trying to calculate the mean rank $\mean{r_i}$ of object $i$ over $R$, which is given by
\begin{eqnarray}
\mean{r_i} &=& \frac{1}{|R|}(r^a_i+r^b_i+\cdots) \nonumber \\
           &=& \frac{1}{|R|C}\bigl(\tilde{s}(r^a_i)+\tilde{s}(r^b_i)+\cdots\bigr).
\end{eqnarray}

Thus we can conclude that calculating $\mean{r_i}$ over $R$ is equivalent to calculating $\mean{s_i}$ over the corresponding rankings in the \Ns~of $R$ under the mapping. See Fig.~\ref{mapping}(c). We will be working exclusively in the \Ns~from now on.

\subsection{Statistical Mechanical Formulation}
It is well known that the canonical ensemble formalism greatly facilitates the calculation of physical quantities in a statistical mechanical system such as that of gaseous particles, while giving identical predictions as the microcanonical ensemble formalism.  The essence of the canonical ensemble formalism lies in one being flexible on which microstates to include in evaluating the partition function; typically, all states of different energies $E$ are allowed, whereas in the microcanonical formalism only states with a specific energy are allowed~\cite{Callen:1985wl}.

We will formulate our problem in an entirely analogous fashion: we include all microstates (rankings) of any number of violations (i.e., not just the MVRs) to form an ensemble.   As usual, the first step is to assign the probability $P(\bds)$ to each ranking $\bds$.  The fairest, most unbiased way to do this is to maximize the entropy

\begin{eqnarray}
\cS = -\sum_{\bds}P(\bds)\ln P(\bds),
\label{entropy}
\end{eqnarray}
subject to certain constraints given by the mean values of the observables~\cite{Jaynes:1957ox,Jaynes:2003ba}.  Our observables are the objects' ranks $\set{s_1,s_2,\ldots}$ and the number of violations $V(\bds)$, which can be written as
\begin{eqnarray}
V(\bds) = \sum_{ij}\Theta (s_j-s_i)\cA_{ij} \equiv \sum_{ij}\Ta\cA_{ij},
\label{defV}
\end{eqnarray}
where $\Theta(s_j-s_i)\equiv\Ta$ is the Heaviside stepfunction.  Therefore, the constraints are given as
\begin{eqnarray}
\mean{s_i} = \sum_{\bds}s_iP(\bds)\text{ and }
\mean{V} = \sum_{\bds}V(\bds)P(\bds).
\label{constraints}
\end{eqnarray}

Using the Lagrange multiplier method on $\cS$, we get the following solution for $P(\bds)$:

\begin{eqnarray}
P(\bds) =\frac{\ex{\bda\cdot\bds+\beta V(\bds)}}{Z} \equiv \frac{\ex{-(H_0(\bds)+H_1(\bds))}}{Z},% = \frac{\ex{-H(\bds)}}{Z},
\label{bmfactor}
\end{eqnarray}
where $\bda=\set{\alpha_1,\ldots,\alpha_n}$ and $\beta$ are the Lagrange multipliers, and $H_0\equiv-\bda\cdot\bds=-\sum_i\alpha_is_i$ and $H_2\equiv-\beta V$ are called Hamiltonians.  The normalizing factor $Z$ is the usual partition function, given by
\begin{eqnarray}
Z = \sum_{\bds} P(\bds) = \sum_{\bds}\ex{-(H_0(\bds)+H_1(\bds))}.
\label{partitionZ}
\end{eqnarray} 

Since we stipulated that all rankings with the same value of $V$ be given equal probability, $P(\bds)$ of Eq.~\eref{bmfactor} can be a function of $V$ only.  Therefore, all values of $\alpha$ are actually equal to $0$.  They are formally retained in Eq.~\eref{bmfactor} because they will serve as the differentiation variables when calculating $\mean{s_i}$ from Eq.~\eref{constraints}:

\begin{eqnarray}
\mean{s_i} = \sum_{\bds} s_iP(\bds) = \frac{1}{Z}\frac{\partial Z}{\partial\alpha_i}\biggr|_{\vec{\alpha}=0}=-\frac{\partial F}{\partial \alpha_i}\biggr|_{\vec{\alpha}=0},
\label{defmeansi}
\end{eqnarray}
where $F=-\ln Z$ is the free energy.  On the other hand, $\beta$ will be a negative real number, as we are trying to minimize (i.e. discourage the formation of) violations.  Furthermore, the MVR limit is given by $\beta=-\infty$.

The model is now fully defined. Before we proceed, let us briefly discuss the advantages of having this theoretical model and solutions to it.  The first advantage, as mentioned earlier, is that our formulation guarantees a fair sampling of the phase space \emph{by definition}, whereas exact or heuristic methods merely attempt to find one specific MVR of a paired comparison.  The second advantage is that, although Eq.~\eref{bmfactor} lends itself to Monte Carlo simulations in a straightforward manner (if it can easily be done for $\beta=-\infty$ at all) for numerical results, having an analytical solution by which we can calculate $\mean{\bds}$ for any paired comparison would be more convenient, possibly requiring no simulation effort at all.

\section{Perturbation Expansion}
We are not currently aware of a general, closed-form solution for Eq.~\eref{partitionZ}, and therefore for Eq.~\eref{defmeansi}.  However, in statistical physics there are a number of sophisticated approximation methods available, such as the perturbation expansion technique which forms the basis of the solution developed here.

First, we re-express the partition function $Z$:
\begin{eqnarray}
Z = \sum_{\bds}\e^{-(H_0+H_1)} &=& Z_0\sum_{\bds}\frac{\e^{-H_0}}{Z_0}\e^{-H_1} \nonumber \\
&=& Z_0\sum_{\bds}P_0(\bds)\e^{-H_1} = Z_0\av{\e^{-H_1}},\nonumber\\
\label{defZ}
\end{eqnarray}
where $Z_0=\sum_{\bds}\e^{-H_0}$, and $\av{\cdots}$ denotes the average of a quantity in the ensemble of $H_0$, in which ranking $\bds$ is given a probability $P_0(\bds)=\e^{-H_0}/Z_0=\e^{\vec{\alpha}\cdot\bds}/Z_0$.  Thus we are treating $H_1$ as the perturbation to $H_0$.

$H_0$ has a very simple meaning: object $i$ is ranked at $s_i$ with probability $P_i(s_i) \propto \ex{\alpha_is_i}$, independent of any other object.  This is also reflected in the fact that $Z_0$ is a product of independent functions of $\alpha_i$:
\begin{eqnarray}
Z_0 = \sum_{\bds}\ex{-H_0(\bds)}&=& \sum_{\bds}\ex{\sum_i\alpha_is_i} %&=& \sum_{\bds}\prod_{i=1}^n\ex{\alpha_is_i} \nonumber \\
= \prod_{i=1}^n\sum_{s_i=1}^N\ex{\alpha_is_i} \nonumber \\
&=& \prod_{i=1}^n\frac{\ex{\alpha_i}(1-\ex{N\alpha_i})}{1-\ex{\alpha_i}} \equiv \prod_{i=1}^n z(\alpha_i).
\label{Z0}
\end{eqnarray}

Therefore, we have
\begin{eqnarray}
P_i(s_i) = \frac{\ex{\alpha_is_i}}{z(\alpha_i)} = \frac{\ex{\alpha_i(s_i-1)}(1-\ex{\alpha_i})}{1-\ex{N\alpha_i}}.
\label{pisi}
\end{eqnarray}

Now, with $H_1=-\beta V$, $Z$ and $F$ become 
\begin{subequations}
\label{ZandF}
\begin{eqnarray}
Z &=& Z_0\av{\ex{\beta V}} = Z_0\biggl(1+\sum_{l=1}^{\infty}\frac{\beta^l}{l!}\av{V^l}\biggr),\label{Znew} \\
F &=& -\ln Z_0-\ln\biggl(1+\sum_{l=1}^{\infty}\frac{\beta^l}{l!}\av{V^l}\biggr) \equiv F_0+F_1. \label{Fnew} \nonumber \\
\end{eqnarray}
\end{subequations}

Therefore, Eq.~\eref{defmeansi} can be written as
\begin{eqnarray}
\mean{s_i} &=& -\biggl[\frac{\partial F_0}{\partial \alpha_i} + \frac{\partial F_1}{\partial \alpha_i}\biggr]\biggr|_{\vec{\alpha}=0} \nonumber \\
&=& \frac{1}{z(\alpha_i)}\frac{\partial z(\alpha_i)}{\partial\alpha_i}\biggr|_{\alpha_i=0} - \frac{\partial F_1}{\partial \alpha_i}\biggr|_{\vec{\alpha}=0} \equiv \frac{N+1}{2} + \Delta_i. \nonumber \\
\label{defmeansi2}
\end{eqnarray}
We can understand this intuitively. The first term $(N+1)/2$ comes from $H_0$, in which an object can assume any rank among $1,\ldots,N$ with uniform probability $1/N$ (with $\alpha=0$).  $(N+1)/2$ is simply its mean, the midpoint.  The second term $\Delta_i$ is, then, the ``shift'' from this midpoint, caused by correlations with other objects via $V$.

\subsection{Evaluation of $\av{V^l}$}
Eq.~\eref{ZandF} tells us that in order to evaluate $Z$ or $F$, we need to calculate $\av{V^l} = \av{(\sum_{ij}\Theta_{ij})^l}$ (with a tacit understanding that the summation is done over $(i,j)$ for which $\cA_{ij}=1$).

Let us look at some examples.  First, consider $\av{V}$:
\begin{eqnarray}
\av{V} = \av{\sum_{ij}\Ta} = \sum_{ij}\av{\Ta}.
\end{eqnarray}

Since $\Ta$ is equal to $1$ when $s_i<s_j$ and $0$ otherwise, $\av{\Ta}$ is simply equal to the probability that $s_i<s_j$. We can easily calculate it using Eq.~\eref{pisi}:

\begin{eqnarray}
\av{\Ta} &=& \sum_{s_i<s_j} P_i(s_i)P_j(s_j) = \sum_{s_i=1}^N\sum_{s_j=s_i}^N P_i(s_i)P_j(s_j) \nonumber \\
&=& \frac{1-\ex{\al_i}-\ex{N\al_j}\bigl(1-\ex{\al_i}(\ex{\al_j}+\ex{N\al_i}(1-\ex{\al_j}))\bigr)}{(1-\ex{N \al_i}) (1-\ex{N \al_j}) (1-\ex{\al_i+\al_j})}. \nonumber \\
\label{theta01}
\end{eqnarray}
The validity of this can be checked quickly by noting its limiting value for $\alpha_i,\alpha_j\to0$, which is $\half+\frac{1}{2N}\simeq\half$.  This is as expected, since if we randomly put two teams at any position with uniform probability (this is what $\alpha\to0$ means), there will be $\frac{1}{2}$ chance that one will be ranked higher than the other.

Next, consider $\av{V^2}=\sum_{ij}\sum_{km}\av{\Theta_{ij}\Theta_{km}}$. We have three types of products $\Theta_{ij}\Theta_{km}$, depending on the relation between $(i,j)$ and $(k,m)$:
\begin{enumerate}
\item $(i,j)=(k,m)$ so that $\Theta_{ij}\Theta_{km}=\Theta_{ij}^2=\Theta_{ij}$, since $\Theta$ is either $1$ or $0$.
\item $(i,j)$ and $(k,m)$ share no index, so that $i\ne k,m$ and $j\ne k,m$. We shall call this a \emph{disconnected} product.
\item $(i,j)$ and $(k,m)$ share one index, for example $j=k$ and $i\ne m$. We shall call this a \emph{connected} product.
\end{enumerate}

Terms of the first kind will sum to $\av{V}$.  In the second case, since the positions of $i$ and $j$ are independent of those of $k$ and $m$, we can easily see that $\av{\Theta_{ij}\Theta_{km}}=\av{\Theta_{ij}}\av{\Theta_{km}}$. The third case is more interesting, because $\av{\Theta_{ij}\Theta_{jm}}$ does not factor like the previous case since $\Ta$ and $\Taa$ are ``connected'' via $j$.  We can see this explicitly by calculating its average in a manner similar to Eq.~\eref{theta01}:

\begin{eqnarray}
\av{\Theta_{ij}\Theta_{jm}} &=& \sum_{s_i<s_j<s_m} P_i(s_i) P_j(s_j) P_m(s_m) \nonumber \\
                            &=& \sum_{s_i=1}^{s_j}\sum_{s_j=1}^N\sum_{s_m=s_j}^NP_i(s_i)P_j(s_j)P_m(s_m). \nonumber \\
\label{theta02}
\end{eqnarray}
Its limiting value for $\alpha_i,\alpha_j,\alpha_m\to 0$ is $\frac{1}{6}$ not $\frac{1}{4}$, as it would have been if the product factored. 

Therefore, $\av{V^2}$ can be written, schematically, as
\begin{eqnarray}
\av{V^2} = \av{V}+\sum\underbrace{\Ta\Tb}_{\mathrm{disconnected}}+\sum\underbrace{\Ta\Tb}_{{\mathrm{connected}}}. \nonumber \\
\end{eqnarray}
The reason for stressing the distinction between disconnected and connected products will become clear in later sections.

Here, let us study the general implication of $\T^2=\T$ on $V^l$. Assume that we have applied the rule exhaustively to every term of $V^l$.  Then $V^l$ will have become a linear combination of all possible products of $1,2,\ldots,l$ distinct $\T$s, namely,
\begin{eqnarray}
V^l = \sum_{k=1}^{\infty}a_{lk}\phi_k,
\end{eqnarray}
where $\phi_k$ is the sum of all possible products of $k$ distinct $\T$'s, and $\set{a_{lk}}$ are appropriate integer coefficients. For example, $V^3$ can be written as
\begin{eqnarray}
V^3 &=& (\Theta_{12}+\Theta_{34}+\Theta_{56}+\cdots)^3 \nonumber \\
    &=& 6(\Theta_{12}\Theta_{34}\Theta_{56}+\cdots) \nonumber \\
    & & + 6(\Theta_{12}\Theta_{34}+\Theta_{12}\Theta_{56}+\Theta_{34}\Theta_{56}+\cdots) \nonumber \\
    & & + (\Theta_{12}+\Theta_{34}+\Theta_{56}+\cdots) \nonumber \\
    &=& 6\phi_3+6\phi_2+\phi_1 = a_{3,3}\phi_3+a_{3,2}\phi_2+a_{3,1}\phi_1.\nonumber \\
\end{eqnarray}

The meaning of $a_{lk}$ is clear: it is the number of ways of selecting one $\T$ from each $V$ such that the selection of size $l$ includes $k$ distinct types of $\T$s.  Now, consider a particular set of $k$ distinct $\T$s. The number of ways of choosing $l$ elements from the set, without the constraint that each of $k$ types must be present at least once, is simply $k\times k\times\cdots = k^l$.  This is also the sum of the number of ways that the selection contains $1,2,\ldots,k$ distinct types of $\T$ from the set.  This leads to the relation $\sum_{m=1}^k{k\choose m}a_{lm}=k^l$ or, equivalently,
\begin{eqnarray}
a_{l,k} = k^l-\sum_{m=1}^{k-1}{k\choose m}a_{l,m}.
\end{eqnarray}
Multiplying each side by $z^l/l!$ and summing them over $l=1,2,\ldots,\infty$ gives the recursion relation for its generating function $g_k(x)=\sum_{l=1}^{\infty}x^la_{lk}/l!$:
\begin{eqnarray}
g_k(z) = (\ex{kz}-1)-\sum_{m=1}^{k-1}{k\choose m}g_m(z).
\end{eqnarray}
This implies $g_k(z)=(\ex{z}-1)^k$ by mathematical induction, with the initial condition $g_1(z)=\sum_{l=1}^{\infty}z^la_{l,1}/k!=(\ex{z}-1)$ and $a_{l1}=1$.  This is the solution to the same result previously reported by us in~\cite{Park:2005qi}.

Since $\phi_1=V$, $Z$ of Eq.~\eref{Znew} now becomes
\begin{eqnarray}
\frac{Z}{Z_0} &=& 1+\sum_{l=1}^{\infty}\frac{\beta^l}{l!}\av{V^l}
               = 1+\sum_{l=1}^{\infty}\frac{\beta^l}{l!}\biggl[\sum_{k=1}^{\infty}a_{lk}\av{\phi_k}\biggr] \nonumber \\
              &=& 1+\sum_{k=1}^{\infty}\biggl[\sum_{l=1}^{\infty}\frac{\beta^l}{l!}a_{lk}\biggr]\av{\phi_k} = 1+\sum_{k=1}^{\infty}(\ex{\beta}-1)^k\av{\phi_k} \nonumber \\
              &\equiv& 1+\sum_{l=1}^{\infty}\eta^l\av{\phi_l}. 
\label{perturbZ2}
\end{eqnarray}
This looks as if we have replaced $\av{V^l}$ and $\beta$ in Eq.~\eref{ZandF} with $\av{\phi_l}$ and $\eta\equiv(\ex{\beta}-1)$.  But what we have here is more than a facelift: thanks to $\eta$, the evaluation of $\av{\phi_l}$ up to any finite $l$ gives us an \emph{infinite}-order expansion in terms of $\beta$.  This is analogous to the ``partial resummation'' in field theory.  The MVR limit under this change of variables is $\eta=-1$.

\subsection{Expansion of the Free Energy}
Using Eq.~\eref{perturbZ2}, $F_1$ of Eq.~\eref{Fnew} can also be written as a power series of $\eta$:
\begin{eqnarray}
F_1 = -\ln\biggl(1+\sum_{l=1}^{\infty}\eta^l\av{\phi_l}\biggr)= -\sum_{l=1}^{\infty}\eta^l f_l,
\label{F_1}
\end{eqnarray}

where its coefficients $f_l$ can be determined by the Taylor series
\begin{eqnarray}
-\ln(1+x) = - x + \frac{x^2}{2} - \frac{x^3}{3} + \cdots.
\label{Taylorln}
\end{eqnarray}

For example, the first three $f$'s are
\begin{eqnarray}
f_1 &=& \av{\phi_1}, \nonumber \\
f_2 &=& \av{\phi_2}-\frac{\av{\phi_1}^2}{2}, \nonumber \\
f_3 &=& \av{\phi_3}-\av{\phi_2}\av{\phi_1}+\frac{\av{\phi_1}^3}{3}.\label{cumulants}
\end{eqnarray}

Since $\phi_l$ is a sum of products of $\T$s, this tell us that $F_1$ is a linear combination of products of their unperturbed averages.  For instance, it has terms made up of powers of $\av{\Ta}$: i.e., $\av{\Ta}$ from $\av{\phi_1}$, $\av{\Ta}^2$ from $\av{\phi_1}^2$, and so forth, which all add up to a quantity $\psi[\Ta]$. We will call this quantity the \emph{contribution} of $\Ta$ to the free energy Eq.~\eref{F_1}:
\begin{eqnarray}
\psi[\Ta] = -\eta\av{\Ta} + \frac{\eta^2}{2}\av{\Ta}^2-\frac{\eta^3}{3}\av{\Ta}^3+\cdots.
\label{psi1}
\end{eqnarray}

Similarly, we can define $\psi[\Ta\Tb] (\Ta\ne\Tb)$, which is the sum of the products of averages of products involving both $\Ta$ and $\Tb$. These are trickier than Eq.~\eref{psi1}: we have $\av{\Ta\Tb}$ from $\av{\phi_2}$, $\av{\Ta}\av{\Tb}$ from $\av{\phi_1}^2/2$ in $f_2$, $\av{\Ta\Tb}\av{\Ta}$ from $\av{\phi_2}\av{\phi_1}$ in $f_3$, and so forth.  These add up to give
\begin{eqnarray}
\psi[\Ta\Tb] &=& -\eta^2(\av{\Ta\Tb}-\av{\Ta}\av{\Tb}) \nonumber \\
&+& \eta^3(\av{\Ta\Tb}-\av{\Ta}\av{\Tb})(\av{\Ta+\Tb}) \nonumber \\
&+& \cdots.
\label{psi2}
\end{eqnarray}

\begin{figure}[t]
\includegraphics[width=60mm]{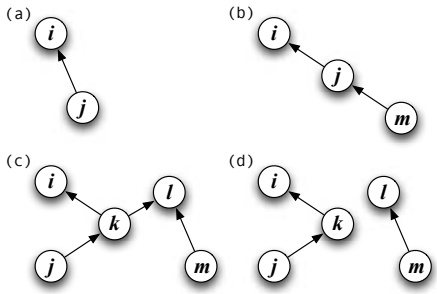}
\caption{Diagrams representing $\Theta$ and their products. (a) $\Theta_{ij}$ is visualized as an arrow pointing to $i$ from $j$. Products of multiple $\T$s are drawn as juxtaposition of arrows. (b) A connected product $\Ta\Taa$ is drawn as a connected diagram. (c) A connected diagram representing $\Theta_{ik}\Theta_{kj}\Theta_{lk}\Theta_{lm}$. (d) A disconnected diagram representing $\Theta_{ik}\Theta_{kj}\Theta_{lm}$.}
\label{visrep}
\end{figure}

Before we proceed any further, let us introduce the diagrammatic representation of products of $\T$s, shown in Fig.~\ref{visrep}.  Here, a single $\Ta$ is drawn as an arrow pointing to $i$ from $j$ (Fig.~\ref{visrep}(a)).  A juxtaposition of two or more arrows will be understood as a product of the corresponding $\T$s, such as Fig.~\ref{visrep}(b) which stands for $\Ta\Taa$.  Note that the diagram for the connected product $\Ta\Taa$ consists of a single component.  In graph theory parlance, such a diagram is also called \emph{connected}.  As a matter of fact, a connected product implies a connected diagram and vice versa in this representation scheme.   Examples of connected and disconnected diagrams (i.e., products) of more than two $\T$s are shown in Fig.~\ref{visrep}(c) and (d).

Therefore, evaluating the free energy $F_1$ means calculating $\psi$ of all possible diagrams, and Eqs.~\eref{psi1} and \eref{psi2} are such calculated up to $\Ord(\eta^3)$.  
However, any power of $\av{\Ta}$ can only come from Taylor-expanding the logarithm of $\eta\av{\Ta}$ contained in the $\eta\av{\phi_1}$ of Eq.~\eref{perturbZ2}, we can easily see that $\psi[\Ta]$ has to be
\begin{eqnarray}
\psi[\Ta] = -\ln\bigl(1+\eta\av{\Ta}\bigr).
\label{psi1-1}
\end{eqnarray}
Taylor-expanding this to $\Ord(\eta^3)$ indeed gives Eq.~\eref{psi1}.

The situation is similar for $\psi[\Ta\Tb]$.  It could have only come from Taylor-expanding $-\ln\bigl(1+\eta\av{\Ta+\Tb}+\eta^2\av{\Ta\Tb}\bigr)$ of Eq.~\eref{F_1}, but this also produces powers solely of $\av{\Ta}$ or $\av{\Tb}$, which we now know are $\psi[\Ta]$ and $\psi[\Tb]$. Therefore, we have
\begin{eqnarray}
\psi[\Ta\Tb] &=& -\ln\bigl(1+\eta\av{\Ta+\Tb}+\eta^2\av{\Ta\Tb}\bigr)\nonumber \\
             & & -\psi[\Ta]-\psi[\Tb] \nonumber \\
             &=& -\ln\biggl[\frac{1+\eta\av{\Ta+\Tb}+\eta^2\av{\Ta\Tb}}{(1+\eta\av{\Ta})(1+\eta\av{\Tb})}\biggr].\nonumber \\
\label{psi2-1}
\end{eqnarray}

Now, assume that $\Ta\Tb$ is disconnected, i.e. $\av{\Ta\Tb}=\av{\Ta}\av{\Tb}$.  Then $\psi[\Ta\Tb]$ vanishes.  This means that only connected products of two $\T$s contribute to the free energy. In fact, this is generally true for products of any number of $\T$s: only connected products (diagrams) make nonzero contributions to the free energy. Therefore, Fig.~\ref{visrep}(c) contributes; Fig.~\ref{visrep}(d) does not.

Now we have an iterative prescription for calculating $\psi$ of any  diagram: from the negative of the logarithm of $1$ plus the terms that represent itself and its subgraphs in Eq.~\eref{perturbZ2}, subtract the $\psi$'s of the subgraphs.  As an example, let us use this rule to evaluate $\psi[\Ta\Tb\Tc]$. The subgraphs of $\Ta\Tb\Tc$ are $\Ta$, $\Tb$, $\Tc$, $\Ta\Tb$, $\Tb\Tc$ and $\Tc\Ta$.  Therefore,  
\begin{eqnarray}
&\psi&[\Ta\Tb\Tc] \nonumber \\
&=&  - \ln\bigl(1+\eta\av{\Ta+\Tb+\Tc}+ \eta^2\av{\Ta\Tb\nonumber\\  & & +\Tb\Tc+\Tc\Ta} + \eta^3\av{\Ta\Tb\Tc}\bigr)\nonumber\\
& &  - \cF[\Ta\Tb] - \cF[\Tb\Tc] - \cF[\Tc\Ta] \nonumber \\
                & & - \cF[\Ta] - \cF[\Tb] - \cF[\Tc].
\end{eqnarray}
Again, this term vanishes unless $\Ta\Tb\Tc$ is connected. 

Our strategy is now clear: identify all connected diagrams (composed of up to some desired number of $\T$), add up their contributions $\psi$, and perform the differentiation of Eq.~\eref{defmeansi2} to obtain $\mean{\bds}$.

\section{Third-Order Expansion}
\begin{figure}[t]
\includegraphics[width=80mm]{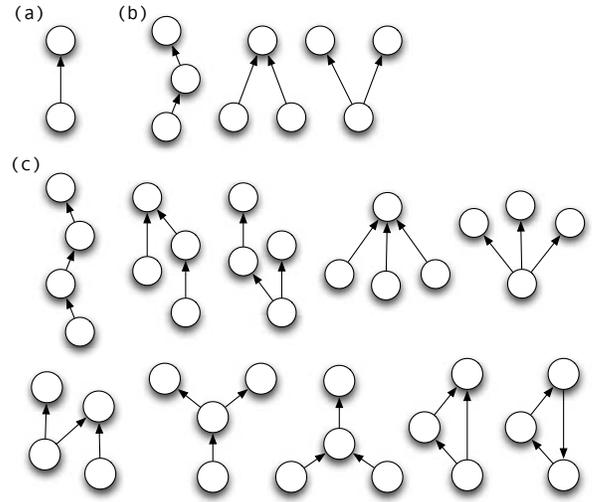}
\caption{Connected diagrams composed of up to three arrows.}
\label{3-diagrams}
\end{figure}

Following the strategy stated in the previous section, we perform a third-order evaluation of $F_1$.  Here, ``third order'' means evaluating $\psi$ for all connected products of up to three $\T$s. (Remember that $F_1$ itself is of infinite order in $\eta$ and $\beta$).  Their diagrams are shown in Fig.~\ref{3-diagrams}.

Since Eq.~\eref{defmeansi2} involves differentiation with respect to a single $\alpha_i$, we realize that a practical method to evaluate averages of the diagrams in ways similar to Eqs.~\eref{theta01} and \eref{theta02} is to first set all $\alpha$'s to $0$ except for $\alpha_i$.  This reduces all $P_j(s_j)$ to $1/N$ except for $P_i(s_i)$, simplifying the calculation.  Now we get two functions of $\alpha_i$ for Fig.~\ref{3-diagrams}(a) -- one for $i$ at the top and another for $i$ at the bottom --- for instance.  We then multiply each by the number of such diagrams present in the data set (the total ``losses'' and ``wins'' of the object $i$) to obtain the total shift of the diagram to $\mean{s_i}$.  For example, the shift $\Delta^{\mathrm{loss}}$ due to a loss by $i$ is
\begin{eqnarray}
\Delta^{\mathrm{loss}} &=& -\frac{\partial\psi[\Ta]}{\partial\alpha_i}\biggr|_{\alpha_i,\alpha_j=0} = \frac{\partial\ln(1+\eta\av{\Ta})}{\partial\alpha_i}\biggr|_{\alpha_i,\alpha_j=0} \nonumber \\
&=& \frac{1}{1+\eta\av{\Ta}}\frac{\partial\av{\Ta}}{\partial\alpha_i}\biggr|_{\alpha_i,\alpha_j=0} = -\frac{(N^2-1)\eta}{6(2N+(N+1)\eta)} \nonumber \\
&=& -\frac{N\eta}{6(2+\eta)} \stackrel{\eta=-1}{\longrightarrow} \frac{N}{12}.
\label{shift01}
\end{eqnarray}

This has a positive value for $\eta<0\mbox{}(\beta<0)$, which is anticipated since we expect a loss to increase one's mean rank.  In the MVR limit of $\eta=-1 (\beta=-\infty)$, it reduces to $N/12$.  This is rather a neat result:  if an object $i$ has $L_i$ losses, its mean rank is shifted upwards by $L_i\times N/12$.  By symmetry, a win by $i$ would decrease its mean rank by $N/12$.

Performing similar calculations for higher-order diagrams gives us functions of $\eta$ not unlike that of Eq.~\eref{shift01}, which we do not show here.  We can, however, easily guess them in a couple of cases:  the first diagram of Fig.~\ref{3-diagrams}(b) cannot shift the mean rank of the node in the middle, since it is being ``pushed'' equally up and down.  The same goes for all nodes in the final diagram of Fig.~\ref{3-diagrams}(c), and our calculations indeed give $0$ for both cases.

Next, we will apply the results obtained here to several examples of paired comparisons.

\subsection*{Example 1: Simple Tree}

\begin{table}
\begin{eqnarray}
\begin{array}{llll}
 0 & 0 & 0 & 0 \\
 1 & 0 & 0 & 0 \\
 0 & 1 & 0 & 0 \\
 0 & 1 & 0 & 0 \\
\end{array}
\end{eqnarray}
\caption{A simple binary comparison dataset.}
\label{matrix01}
\end{table}

\begin{figure}[t]
\includegraphics[width=50mm]{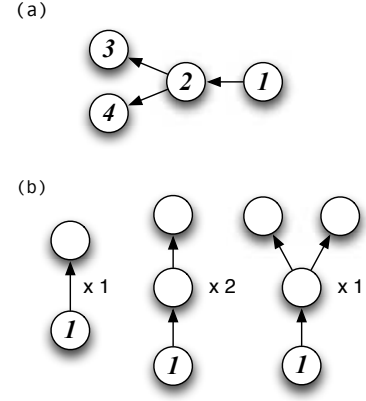}
\caption{(a) The visualization of Table~\ref{matrix01}. (b) connected subgraphs involving object 1.}
\label{example01}
\end{figure}

\begin{figure}[t]
\includegraphics[width=80mm]{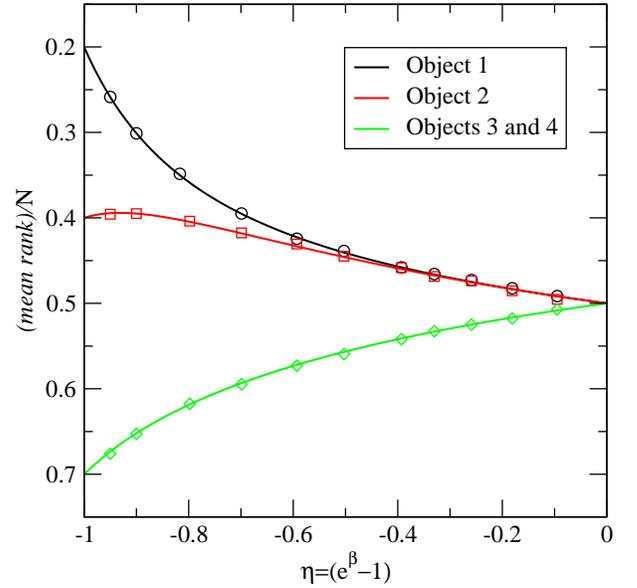}
\caption{The normalized rankings of objects depicted in Fig.~\ref{example01}(a). The perturbation expansion (solid lines) is exact in this case, and agrees perfectly with Monte Carlo simulation results (shapes).}
\label{example1sim}
\end{figure}

As the first example, take a simple comparison whose adjacency matrix is given in Table~\ref{matrix01}. It can be visualized as Fig.~\ref{example01}(a).  We can intuitively see that object $1$ must be ranked at the top, followed by $2$, while $3$ and $4$ should be tied (since they are topologically indistinguishable) at last to obtain a minimum number of violations ($0$).  In order to use our expansion results, we must count the connected subgraphs involving each object, which are shown in Fig.~\ref{example01}(b) for object $1$.  After repeating this process for all objects, we obtain the following solutions for $\mean{\bds}$:

\begin{eqnarray}
\frac{\mean{s_1}(\eta)}{N} &=& \frac{4 \eta ^3+25 \eta ^2+50 \eta +30}{5 (\eta+2) \left(\eta ^2+6 \eta +6\right)} \nonumber \\
\frac{\mean{s_2}(\eta)}{N} &=&\frac{3 \eta ^3+25 \eta ^2+50 \eta +30}{5 (\eta+2) \left(\eta ^2+6 \eta +6\right)} \nonumber \\
\frac{\mean{s_3}(\eta)}{N} &=& \frac{3 \eta ^3+30 \eta ^2+80 \eta +60}{10 (\eta+2) \left(\eta ^2+6 \eta +6\right)} = \frac{\mean{s_4}(\eta)}{N}.
\label{treesoln}
\end{eqnarray}
Since we have considered all subgraphs appearing in Fig.~\ref{example01}(a), these must be exact.  An excellent agreement with Monte Carlo simulation results for the entire range of $\eta$ in Fig.~\ref{example1sim} confirms this.  

A closer inspection of Fig.~\ref{example1sim} tells us that the mean ranks are $(N+1)/2=N/2$ for $\eta=0(\beta=0)$ as expected, while $\mean{s_1}$ and $\mean{s_2}$ begin to diverge noticeably around $\eta=-0.6 (\beta=-0.9)$.  Since objects $1$ and $2$ are indistinguishable on the first order (they both have a win-loss differential of $1$), this has to be the point where second and third-order diagrams begin to contribute significantly.  The MVR solutions at $\eta=-1$ also agree with Eq.~\eref{scaling}.

Larger data sets will certainly contain fourth- or higher-order connected diagrams, and therefore our third-order expansion will be an approximation.  However, it is still interesting to see how valid our third-order expansion can be as a true approximation.  We study this issue with the following examples.

\subsection*{Example 2: 5-Player Round-Robin Tournament}

\begin{table}
\begin{eqnarray}
\begin{array}{lllll}
 0 & 0 & 0 & 0 & 0 \\
 1 & 0 & 0 & 0 & 0 \\
 1 & 1 & 0 & 0 & 0 \\
 1 & 1 & 1 & 0 & 0 \\
 1 & 1 & 1 & 1 & 0
\end{array}
\end{eqnarray}
\caption{The adjacency matrix for the tournament 5 in~\cite{Ali:1986ey}.}
\label{rrobin}
\end{table}

\begin{figure}[t]
\includegraphics[width=40mm]{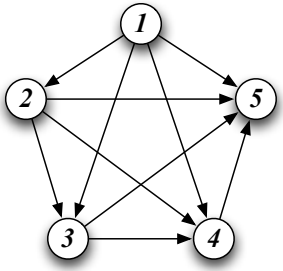}
\caption{The visualization of Table~\ref{rrobin}.}
\label{tour5diagram}
\end{figure}

\begin{figure}[t]
\includegraphics[width=80mm]{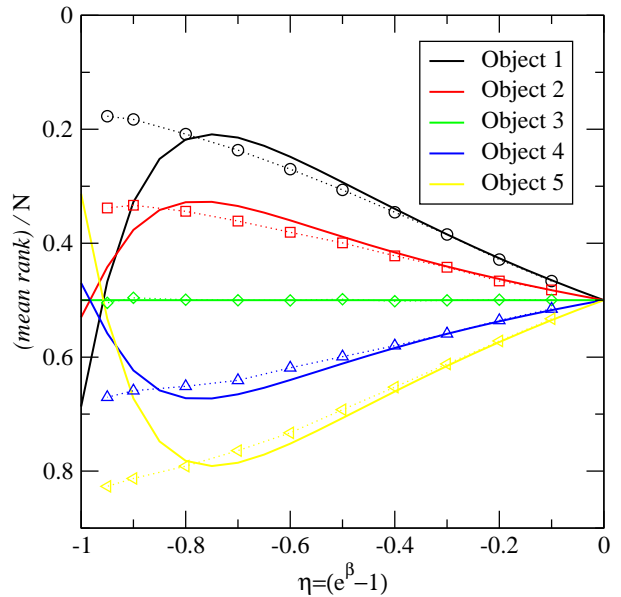}
\caption{The third-order perturbation expansion (solid lines) and Monte Carlo simulation results (shapes) for the tournament in Table~\ref{rrobin}.  There is a wide range of $\eta$ for which the perturbation gives a correct prediction of standings, followed by a region of haphazard behavior near $\eta=-1$ caused by missing higher-order diagrams.}
\label{tour5}
\end{figure}

A \defn{round-robin tournament} is a complete paired comparison.  Ali~\etal~studied the minimum violations ranking problem of the $5$-player round-robin tournament (with ${5\choose 2}=10$ comparisons) using a couple of heuristic methods on the nine distinct configurations they had identified~\cite{Ali:1986ey}.  One of those tournaments (labeled $5$ in their paper) is given in Table~\ref{rrobin}, and visualized in Fig.~\ref{tour5diagram}. The correct MVR standings should be $\set{1,2,3,4,5}$.

Fig.~\ref{tour5} shows the perturbation and Monte Carlo simulation results.  We see that there exists a wide range of $\eta$ (between $-0.8$ and $0$) in which the perturbation agrees with the simulation well within $2\%$ of $N$ and predicts the relative standings correctly, whereas closer to $\eta=-1$ it behaves in a haphazard way and gives incorrect standings.  This latter behavior is certainly due to the missing higher-order diagrams.  We find almost identical behaviors for the eight other tournaments as well: the perturbation solution is stable and predicts correct standings including ties for a wide range of $\eta$, while near $\eta=-1$ it fails.

Compare this to the heuristic methods of Ali~\etal, which merely succeeds in finding one specific MVR for each the tournament.  Furthermore, strangely enough, they regard the emergence of ``tied ranks'' as an undesirable feature of a method in their study.  We believe to the contrary: it is only fair to give tied ranks to topologically indistinguishable players, which our method does explicitly.

\subsection*{Example 3: American Collegiate Football}
As a final example, we consider a large-scale data set mentioned earlier, American collegiate football. In the year $2004$, $117$ universities played a total of $622$ games in the regular season.  Ranking of colleges based on the outcomes of those  games is of particular interest, because schools ranked high in the ``official'' ranking (currently determined by a combination of human polls and several computer algorithms) get invited to play in prestigious post-season bowl games, while the top two schools get to play in the national championship game~\footnote{\texttt{www.bcsfootball.org}}.  An invitation to these games is not only honorable, but brings practical benefits to the schools, including much publicity through the media, financial compensations worth millions of US dollars, and an increase in applications from prospective students~\cite{Dunnavant:2004ns}.

Due to the small number of games ($\sim11$) a school plays in a season compared to the number of possible opponents ($\sim120$), a simple win-loss differential is a hopelessly insufficient measure of a team's strength. Therefore, the official computer algorithms as well as the hundreds of rankings schemes devised by the fans of the sport employ various criteria that their inventors or developers deem necessary and important in order to assess the college teams' strengths~\cite{Park:2005fb,Keener:1993jj,Harville:2003uj,Callagan:2004mh}. 

One often evaluates the quality of an algorithm by counting the violations in the final ranking it produces~\cite{Martinich:2002dx}. Any MVR will perform the best in this respect by definition, but in a recent study of a particular MVR that was mentioned earlier was shown to have performed as well as or better than other ranking methods in other tests as well~\cite{Coleman:2005vm}.  Again, this study was done for one particular MVR, and so it would be an interesting topic for future research to study how $\mean{\bds}$ from Eq.~\eref{defmeansi2} perform in those tests.

\begin{figure}[t]
\includegraphics[width=80mm]{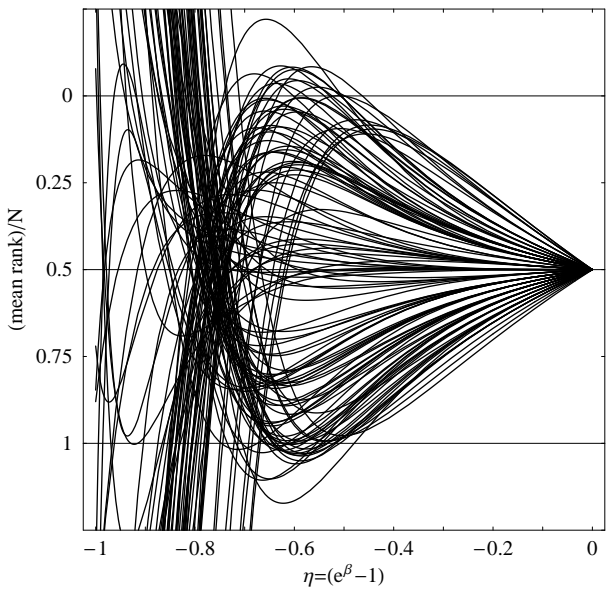}
\caption{The third-order perturbation expansion for $117$ American college football teams in 2004.}
\label{frank}
\end{figure}

In this paper, though, let us concentrate on our third-order perturbation expansion.  In Fig.~\ref{frank} we see that it shows a behavior quite similar to the one we have seen in the previous example: we have a smooth, well-behaved range (though narrower, $-0.3\lesssim\eta<0$), followed by a region of large fluctuations.

\begin{figure}[t]
\includegraphics[width=80mm]{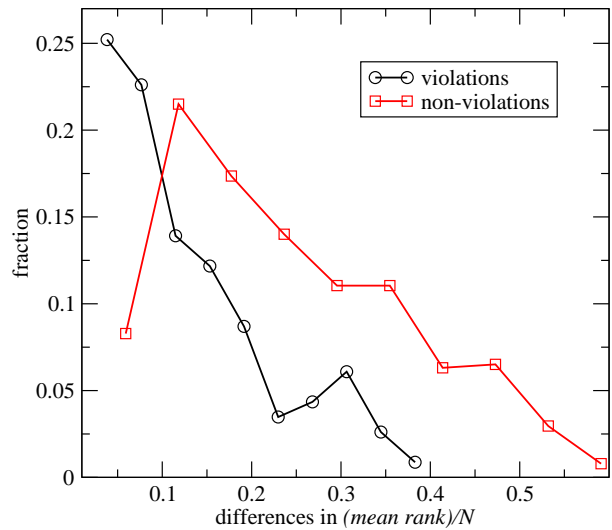}
\caption{The normalized distributions for the differences in $\mean{s}/N$ between object pairs across violations and non-violations, for $\eta=-0.3$. This shows that violations occur between object pairs that are generally closer in their mean ranks.}
\label{viodist}
\end{figure}

In the smooth range, the number of violations is consistently $115$, short of the $53$ found by Coleman.  Nevertheless, it is interesting that the violation decreases rapidly to about a third of $311$ (half the number of games) expected at $\eta=0$ as soon as $\eta$ is tuned away from it (at $\eta=-0.001$), and stays that way over the entire region.  The relative standings of the teams also change very little, suggesting that they may still reflect the correct ones to some degree.  If this is the case, then the violations must be occurring between teams that are closer in $\mean{s}$ so that adding more diagrams are likely to swap these pairs first, causing only limited changes to the standings.  In Fig.~\ref{viodist}, we show the distributions of the separation in $\mean{s}/N$ between the $115$ pairs of teams across violations and the $622-115=507$ pairs that are not, for $\eta=-0.3$.  We clearly see that $\mean{s}/N$ are closer across a violation on average.  Actually, by directly swapping the positions of pairs of schools across a violation in the increasing order of their separation in $\mean{s}/N$, we were able to reduce the number of violations by another $20$ --- $17\%$ of the original $115$ --- down to $95$.  (In order to lower it further, we need to swap objects in a more complicated order. In this sense, a higher-order expansion is precisely what tells us how to do that.)

\section{Summary and Discussion}
In this paper, we have formulated the problem of minimum violations rankings in paired comparisons as a statistical physics model, analogous to the canonical ensemble theory of systems of particles.  This formulation ensured that all configurations with the same number of violations were assigned equal statistical weights in the ensemble, overcoming the conceptual difficulties faced by earlier studies of the problem.

We then developed a diagrammatic perturbation technique for evaluating the free energy $F$, from which we could obtain the mean ranks $\mean{s}$ of each object via differentiation.  Making use of the property $\Theta^2(x-y)=\Theta(x-y)$ of the Heaviside step function, we were able to find an iterative prescription for evaluating infinite-order contributions of finite diagrams (i.e., partial resummation). The fact that we only needed to consider connected diagrams greatly facilitated the expansion, and a third-order perturbation expansion was carried out explicitly.

We were able to use these results on any type of paired comparison data set, having only to count the number of connected subgraphs it contained.  We studied three example data sets.  In the first example, the perturbation expansion had to be exact, and we confirmed this by comparing it to a Monte Carlo simulation.  In the latter two examples we were interested in the performance of the expansion as a true approximation.  We witnessed the existence of a region in which the perturbation expansion was stable and well-behaved. It was followed by large fluctuations near $\eta=-1$ due to the missing higher-order diagrams.  The reliability of this stable, smooth range as the predictor of the true mean minimum violations $\mean{\bds}$ at $\eta=-1$ would depend on how high an order we perform the expansion into, but already on third order we were able to predict the relative standings perfectly for the $5$-player round-robin tournament, and to some reasonable degree for the $117$-player American college football schedule.

We propose several directions for future research based on this work.  First, performing the expansion into higher orders is always an obvious possibility.  Second, developing faster algorithms or programming more efficient codes for counting the connected subgraphs is desirable.  Third, a generalization of the theory to cases where $\cA_{ij}>1$ (i.e., allowing multiple comparisons or considering the weight of a comparison, such as the score difference in a sports match) is allowed would be welcome. We hope that these efforts will prove helpful in the practical application of this method to other large-scale data sets.

Lastly, we note that the entire perturbation technique developed in this paper is a significant improvement over the one introduced earlier in~\cite{Park:2004sm} in the context of the theory of exponential random graphs. We are hoping to see more developments in that area as well.

\begin{acknowledgments}
The author would like to thank Mark Newman, B. Jay Coleman, Elizabeth Leicht, Petter Holme, Gourab Ghoshal, Henry Kim and Willemien Ketz for helpful discussions, and the Santa Fe Complex Systems Summer School 2005 for the opportunity to pursue this interdisciplinary project.  This research was supported by the National Science Foundation under grant number DMS-0405348.
\end{acknowledgments}

\end{document}